\documentclass[%
 reprint,
 amsmath,amssymb,
aps,
]{revtex4-1}

\usepackage{graphicx}
\usepackage{dcolumn}
\usepackage{bm}
\usepackage{epstopdf}
\usepackage{mathtools}
\usepackage{amsmath}
\usepackage{array}
\usepackage{amsmath,amssymb}
\usepackage{lipsum}
\usepackage{subcaption}
\usepackage{CJK}

\begin{document}


\title{Modal Group-Velocity Mismatch Induced Intermodal Modulation Instability in Step-index Fiber}

\author{Partha Mondal}
 \altaffiliation[ ]{Department of Physics, Indian Institute of Technology, Kharagpur - 721302, India, Email: parthaphotonica@gmail.com}
 
\author{Shailendra K. Varshney}%

\altaffiliation[ ]{Department of E $\&$ ECE, Indian Institute of Technology, Kharagpur - 721302, India}

\date{\today}

\begin{abstract}
We present detailed experimental study on noise-seeded intermodal modulation instability (IM-MI) in normal dispersion region of a conventional step-index fiber. The sharp refractive index contrast between core and cladding leads to large group velocity mismatch between the spatial modes, coaxing to efficient IM-MI and generation of multiple spectral peaks along with Raman peaks. Evolution of the spectrum with pump powers and fiber lengths are observed. Experimental findings are well supported with the theoretical framework based on bimodal-MI model considering the distinct dispersion parameters of the participating modes.
\end{abstract}

\maketitle

Modulation instability (MI) is an ubiquitous natural phenomena that leads to spontaneous pattern formation due to stochastic fluctuation in large variety of systems such as water wave instabilities \cite{water_wave1,water_wave2}, surface-waves of sand dunes \cite{sand_wave}, optical system \cite{matter_wave,B_E_condensation} etc. In case of optical fiber, MI occurs as the interplay between dispersion and nonlinearity where a continuous or quasi-continuous wave propagating through nonlinear medium breaks-up into ultrashot pulses in presence of weak noise or small perturbation. MI has been harnessed extensively and employed in myriad of applications such as for supercontinuum generation based on parametric conversion \cite{MI_SC1,MI_SC2}, generation of optical pulse with high repetition rate \cite{pulse_generation_MI} and so on. The first experimental observation of MI was reported in the anomalous dispersion region of single mode fiber \cite{MI_1st}. Extensive amount of work have been carried out to demonstrate MI in the anomalous dispersion region for a single pump \cite{MI_anomolous_singlepump1,MI_anomolous_singlepump2,MI_anomolous_singlepump3}. The first demonstration of MI in the normal dispersion region of a birefringent fiber for a single frequency was reported by Wabnitz \cite{MI_normal_region1}. Later on, many theoretical and experimental studies on MI have been reported in normal dispersion regime for single and double co-propagating beams \cite{MI_normal_region2,MI_normal_doublepump1,MI_normal_doublepump2,MI_normal_doublepump3,MI_normal_doublepump4,MI_normal_doublepump5}. The influence of group velocity mismatch (GVM) on MI was reported by Drummond \textit{et. al.} \cite{MI_GVM} and showed that MI process gradually decreases as GVM decreases and vanish as GVM becomes zero. The detailed investigation of MI for single and dual pump in normal dispersion region has been reported \cite{MI_normal_region3}. Parallelly, pump photons launched into different spatial modes gives rise to intermodal MI (IM-MI). The details on theoretical and numerical analysis of IM-MI in multimode fiber has been described \cite{IM_MI_Guasoni,IM_MI_OC}. More recently, the experimental observation of IM-MI in the normal dispersion region of graded index fiber has been demonstrated where the effect of the GVM and the variation of dispersion parameters between the participating spatial modes have been neglected \cite{IM_MI_greadedindex_fiber}. To the best of our knowledge, the effect of GVM of the interacting modes has not been reported yet in case of IM-MI.

In this work, we mainly focus on the role of GVM of the participating spatial modes  in the generation of noise-seeded IM-MI peaks.  To investigate this, we have carried out detailed theoretical and experimental work  for the realization of IM-MI in the normal dispersion region of a commercially available step-index Corning$^\circledR$LEAF$^\circledR$ fiber (CLF) considering the GVM and distinct dispersion profile of the participating spatial modes. Step-index fiber provides large GVM between the propagating modes which leads to rich IM-MI dynamics compared to graded-index fiber. For experimental investigation, quasi-continuous pump pulses at 1064 nm are launched where the fiber can support $LP_{01}$, $LP_{11}$ and $LP_{02}$ spatial modes. Pump is injected into the fiber such that it can equally excite  $LP_{01}$ and  $LP_{02}$ modes and this gives rise to multiple IM-MI peaks along with the Raman peaks. Evolution of the spectra with varying pump powers and fiber lengths are recorded. Detailed theoretical formalism based on bimodal-MI model are also presented which support which support our experimental observations with very good accuracy.

\section{THEORETICAL APPROACH}
\begin{figure*}[t]
   \centering
   \includegraphics[width=12.5 cm,height=14.5 cm]{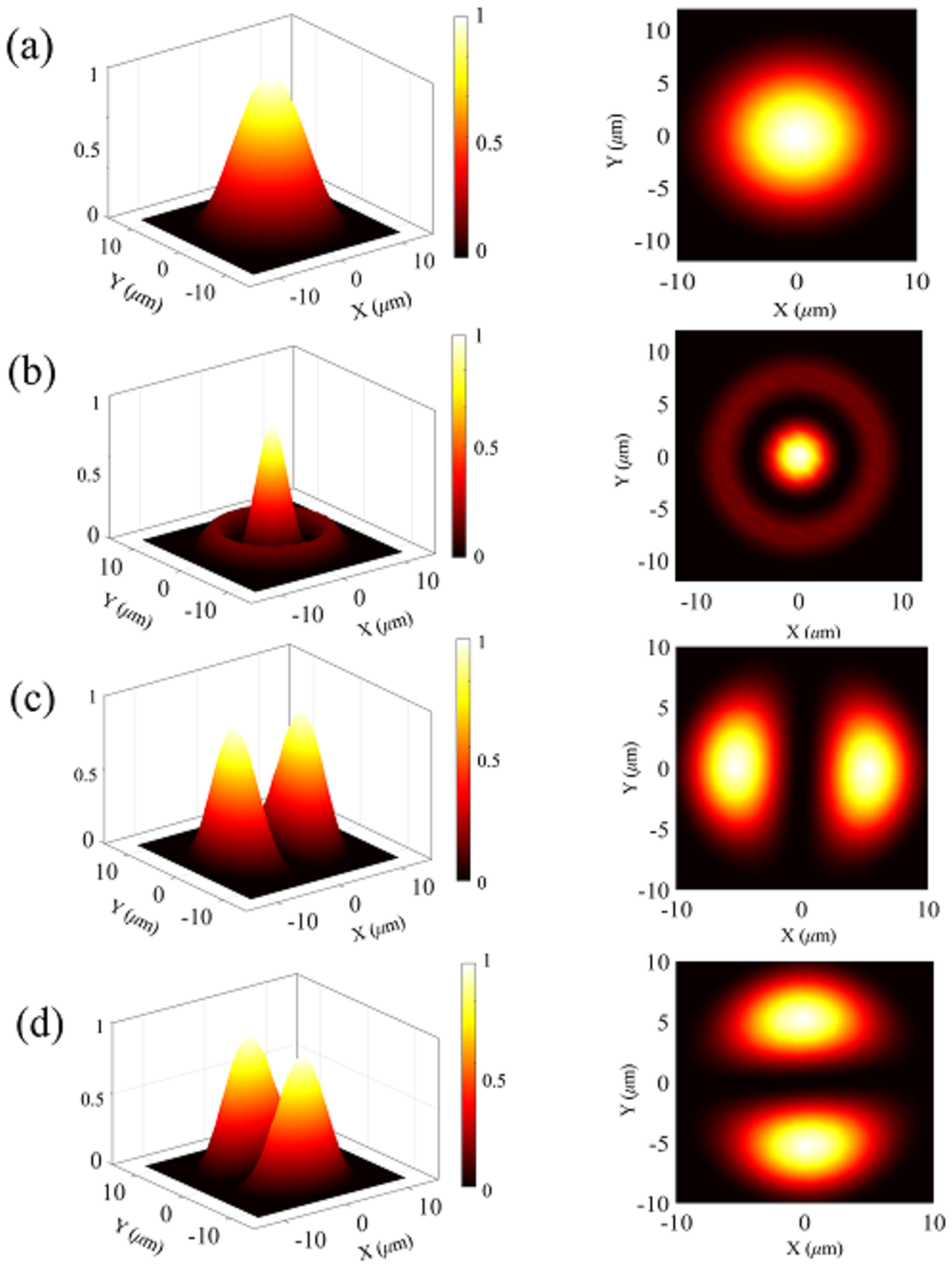}
   \caption{{\small 3D/2D schematic view of the modal profile distribution of (a) $LP_{01}$ (b) $LP_{02}$ (c) $LP_{11x}$ and (b) $LP_{11y}$ at 1064 nm inside CLF.}}
   \label{modal_profile}
   \end{figure*}

\begin{figure}[t]
   \centering
   \includegraphics[width=8 cm,height=4.5 cm]{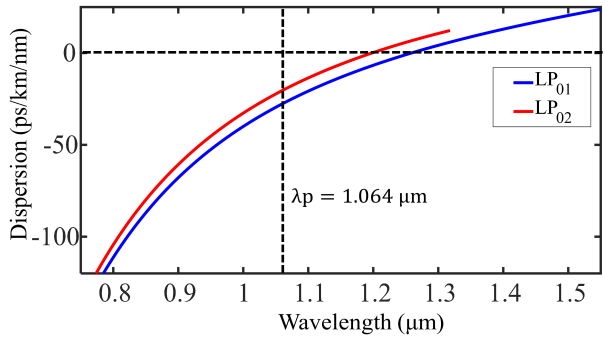}
   \caption{{\small Dispersion characteristics of $LP_{01}$ and $LP_{02}$ modes of CLF.}}
   \label{dispersion_plot}
   \end{figure}

\begin{figure}[t]
   \centering
   \includegraphics[width=8.5 cm,height=4.5 cm]{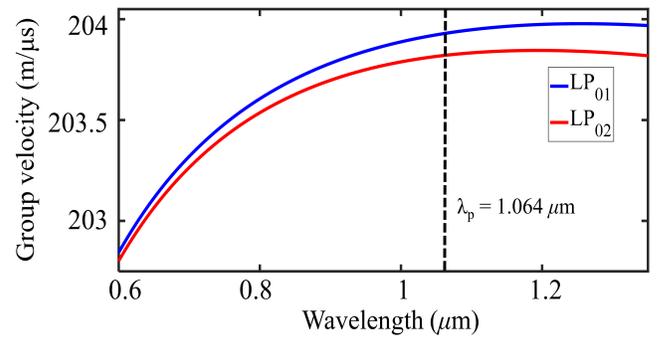}
   \caption{{\small Simulated group velocity for the $LP_{01}$ and $LP_{02}$ spatial modes.}}
   \label{group_velocity}
   \end{figure}

The fiber used in our experiment is a step-index non-zero dispersion shifted CLF which is widely been used in communication. The fiber exhibits effective mode area and numerical aperture (NA) of 72 $\mu m^{2}$ and 0.14, respectively at 1550 nm. Pumping below the cut-off wavelength of the fiber, turns it into a few-mode fiber. At 1064 nm pump wavelength, the supported spatial modes through the fiber are simulated using full-vectorial finite-element method (FEM) based commercial COMSOL software and shown in Fig. \ref{modal_profile}. Simulated results show that the fiber supports $LP_{01}$, $LP_{02}$ and $LP_{11}$ spatial modes at the pump wavelength which are confirmed experimentally and will be discussed later. To demonstrate IM-MI, we excite the pair of circularly symmetric spatial modes $LP_{01}$ and $LP_{02}$ with equal peak power. The dispersion profile of two modes is shown in Fig. \ref{dispersion_plot} and it is observed that the pump at 1064 nm falls in the normal dispersion region for both the modes. The step-index refractive index profile provides distinct group-velocity for different modes as shown in Fig. \ref{group_velocity}. It is observed that the GVM between the two modes increase with the increase in wavelength.  
  \begin{figure}[t]
            \centering
            \includegraphics[width=8.5 cm,height=4.5 cm]{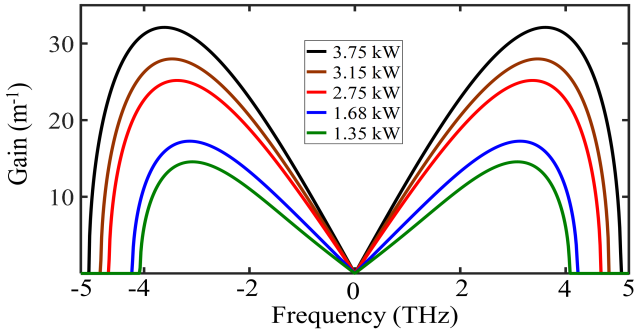}
            \caption{{\small Theoretical gain spectra as a function of frequency shift ($\Omega/2\pi$) for the mode group combination $LP_{01}$ and $LP_{02}$ with identical power in each mode (P=Q).}}
            \label{gain_plot}
            \end{figure}
\begin{figure}[t]
            \centering
            \includegraphics[width=8.5 cm,height=4.5 cm]{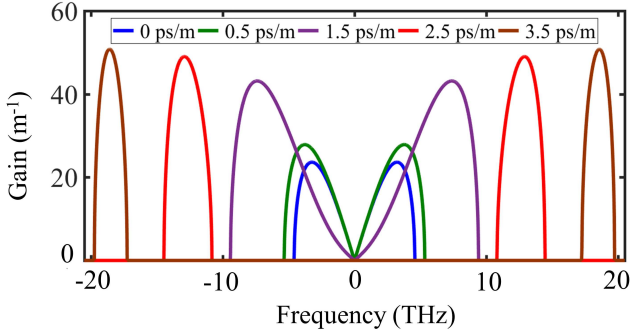}
            \caption{{\small Gain spectra as a function of frequency shift ($\Omega/2\pi$) for different GVM values while the peak power is fixed for the both modes (P = Q = 3.75 kW).}}
            \label{GVM_variation}
            \end{figure}
\begin{figure}[t]
            \centering
            \includegraphics[width=8.5 cm,height=4.5 cm]{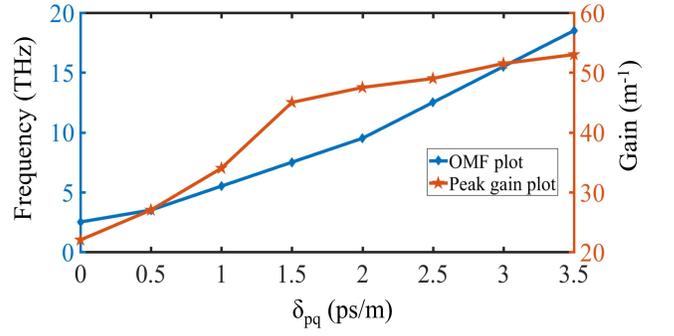}
            \caption{{\small Variation of OMF and peak gain of IM-MI as a function of GVM for fixed pump power (P = Q = 3.75 kW).}}
            \label{OMF_GAIN_shift}
            \end{figure}
    
The nonlinear propagation of the interacting modes with identical carrier frequency $\omega$ satisfies the following set of coupled nonlinear Schrodinger equations (NLSE),
 \begin{equation}
 \begin{aligned}
 \frac{\partial u_{p}}{\partial z}-\frac{\delta_{pq}}{2}\frac{\partial u_{p}}{\partial t}+\frac{i}{2}\beta_{2p}\frac{\partial^{2} u_{p}}{\partial t^{2}}=i\gamma(\textit{f}_{pp}|u_{p}|^{2}\\+2\textit{f}_{pq}|u_{q}|^{2})u_{p}+i\gamma \textit{f}_{pq}u_{q}^{2}u_{p}^{*}exp(2i\Delta\beta z)
 \label{up_equation} 
 \end{aligned}
 \end{equation}
\begin{equation}
 \begin{aligned}
 \frac{\partial u_{q}}{\partial z}+\frac{\delta_{pq}}{2}\frac{\partial u_{q}}{\partial t}+\frac{i}{2}\beta_{2q}\frac{\partial^{2} u_{q}}{\partial t^{2}}=i\gamma(\textit{f}_{qq}|u_{q}|^{2}\\+2\textit{f}_{pq}|u_{p}|^{2})u_{q}+i\gamma \textit{f}_{pq}u_{p}^{2}u_{q}^{*}exp(-2i\Delta\beta z)
 \label{uq_equation} 
 \end{aligned}
 \end{equation}
 where $u_{j}(j=p,q)$ is the field envelopes of the interacting modes, \textit{z} is the propagation distance and \textit{t} is the time. Nonlinear Kerr coefficient, $\gamma$=$\frac{n_{2}\omega}{c}$, c is the velocity of light in vacuum. $\Delta\beta=\beta_{q}-\beta_{p}$, where $\beta_{j}$ is the propagation constant of the corresponding mode, $\beta_{nj}=\frac{\partial^{n}\beta_{j}}{\partial\omega^{n}}$ stands for the nth derivatives of the propagation constant, $\beta_{1j}$ is the inverse of the group velocity, $\beta_{2j}$ is the second order dispersion coefficient. $\delta_{pq}$ indicates the GVM of the participating modes and defined as, $\delta_{pq}=\beta_{1q}-\beta_{1p}$. The overlap function $\textit{f}_{pq}$ is defined as,
 \begin{equation}
  \begin{aligned}
f_{pq}=\dfrac{\int\int |F_{p}(\omega)|^{2}|F_{q}(\omega)|^{2}\,dx\,dy}{[\int\int |F_{p}(\omega)||F_{q}(\omega)|\,dx\,dy]^{2}}
  \label{fpq_equation} 
  \end{aligned}
  \end{equation}
  where, $F_{p}$ and $F_{q}$ are the transverse field distributions of the participating modes. The effective mode area of the two modes can be expressed by $1/f_{pp}$ and $1/f_{qq}$, respectively, where, $1/f_{pq}$ represents the overlap between two interacting modes. We would like to point out that the coherent coupling terms $i\gamma \textit{f}_{pq}u_{q}^{2}u_{p}^{*}exp(2i\Delta\beta z)$ and $i\gamma \textit{f}_{pq}u_{p}^{2}u_{q}^{*}exp(-2i\Delta\beta z)$, in the right hand side of Eqs. (\ref{up_equation}) and (\ref{uq_equation}), depend essentially on the GVM between the spatial modes. For large GVM, these terms are essentially negligible. As, in our case, we use step-index fiber which leads to large GVM between the interacting modes, thus the coherent coupling term is neglected in the preceding calculations. Considering only the incoherent coupling terms, Eqs. (\ref{up_equation}) and (\ref{uq_equation}) can be rewritten as,
   \begin{equation}
   \begin{aligned}
   \frac{\partial u_{p}}{\partial z}-\frac{\delta_{pq}}{2}\frac{\partial u_{p}}{\partial t}+\frac{i}{2}\beta_{2p}\frac{\partial^{2} u_{p}}{\partial t^{2}}=i\gamma(\textit{f}_{pp}|u_{p}|^{2}\\+2\textit{f}_{pq}|u_{q}|^{2})u_{p}
   \label{up_equation_modified} 
   \end{aligned}
   \end{equation}
  \begin{equation}
   \begin{aligned}
   \frac{\partial u_{q}}{\partial z}+\frac{\delta_{pq}}{2}\frac{\partial u_{q}}{\partial t}+\frac{i}{2}\beta_{2q}\frac{\partial^{2} u_{q}}{\partial t^{2}}=i\gamma(\textit{f}_{qq}|u_{q}|^{2}\\+2\textit{f}_{pq}|u_{p}|^{2})u_{q}
   \label{uq_equation_modified} 
   \end{aligned}
   \end{equation}
  To analyze the stability of the steady state solution of Eqs. (\ref{up_equation_modified}) and (\ref{uq_equation_modified}), we introduce a small first-order amplitude and phase perturbation u and v, where,
 \begin{equation}
   \begin{aligned}
   u_{p}=(\sqrt{P}+u)exp[i\gamma(\textit{f}_{pp}P+2\textit{f}_{pq}Q)z]
   \label{up_perturbation} 
   \end{aligned}
   \end{equation}
\begin{equation}
   \begin{aligned}
   u_{q}=(\sqrt{Q}+v)exp[i\gamma(\textit{f}_{qq}Q+2\textit{f}_{pq}P)z]
   \label{uq_perturbation} 
   \end{aligned}
   \end{equation}
Now, we consider perturbation of modulational ansatz with wavenumber K and frequency $\Omega$, of the form,
\begin{equation}
   \begin{aligned}
   u(z,t)=u_{s}(z)exp[i(\Omega t-Kz)]+u_{a}(z)exp[i(-\Omega t+Kz)]
   \label{up_ansaz} 
   \end{aligned}
   \end{equation}
\begin{equation}
   \begin{aligned}
   v(z,t)=v_{s}(z)exp[i(\Omega t-Kz)]+v_{a}(z)exp[i(-\Omega t+Kz)]
   \label{uq_ansaz} 
   \end{aligned}
   \end{equation}
   where, $u_{s}$ and $u_{a}$ represents the amplitude of Stokes and anti-Stokes sidebands for the spatial mode p, respectively, whereas $v_{s}$ and $v_{a}$ corresponds to the spatial mode q. $\Omega$ is the angular offset frequency relative to the pump, $\Omega = \omega - \omega_{p}$, where $\omega_{p}$ is the angular frequency for the pump wavelength. After linearizing Eqs. (\ref{up_equation_modified}) and (\ref{uq_equation_modified}) in u and v and then substituting Eqs. (\ref{up_ansaz}) and (\ref{uq_ansaz}) in it, we arrive at the following eigenvalue equation,
\begin{equation}
   \begin{aligned}
  [M][Y]=K[Y]
   \label{eigenvalue_eq} 
   \end{aligned}
   \end{equation}
   where the eigen vector is defined as, 
\begin{equation}
   \begin{aligned}
  [Y]^{T}=[u_{a},u_{s}^{*},v_{a},v_{s}^{*}]
   \label{eigenvector_eq} 
   \end{aligned}
   \end{equation}
   [M] is the stability matrix of the system defined as,\par

   \vspace{-.5ex}
\begin{frame}

\resizebox{1.05\linewidth}{!}{%
$\displaystyle
M = 
 \begin{bmatrix}
-\dfrac{\Omega\delta_{pq}}{2}+\beta_{2p}\dfrac{\Omega^{2}}{2}+\gamma_{p}\textit{f}_{pp}P & \gamma \textit{f}_{pp}P & 2\gamma \textit{f}_{pq}\sqrt{PQ} & 2\gamma \textit{f}_{pq}\sqrt{PQ} \\
-\gamma \textit{f}_{pp}P & -\dfrac{\Omega\delta_{pq}}{2}-\beta_{2p}\dfrac{\Omega^{2}}{2}-\gamma_{p}\textit{f}_{pp}P&-2\gamma \textit{f}_{pq}\sqrt{PQ} & -2\gamma\textit{ f}_{pq}\sqrt{PQ} \\
2\gamma \textit{f}_{pq}\sqrt{PQ} & 2\gamma f_{pq}\sqrt{PQ} & \dfrac{\Omega\delta_{pq}}{2}+\beta_{2q}\dfrac{\Omega^{2}}{2}+\gamma\textit{f}_{qq}Q & \gamma \textit{f}_{qq}Q \\
-2\gamma \textit{f}_{pq}\sqrt{PQ} & -2\gamma \textit{f}_{pq}\sqrt{PQ}& -\gamma f_{qq}Q & \dfrac{\Omega\delta_{pq}}{2}-\beta_{2q}\dfrac{\Omega^{2}}{2}-\gamma\textit{f}_{qq}Q\\
\end{bmatrix}
$}
\end{frame}

from which we obtain the following dispersion relation, 
\begin{equation}
   \begin{aligned}
det([M]-K[I])=0
   \label{eigenvector_eq} 
   \end{aligned}
   \end{equation}
  \begin{table}
\centering
     \caption{ Calculated IM-MI parameters for different mode combinations at 1064 nm.}
     \begin{tabular}{ccccccc}
     \hline
     $p$ & $LP_{01}$ & $LP_{01}$ & $LP_{02}$ \\
     $q$ & $LP_{11}$ & $LP_{02}$ & $LP_{11}$ \\
     \hline
     $ f_{pp}$ $ (1/\mu m^{2}) $ & $0.0257$ & $0.0257$ & 0.015 \\
      $ f_{qq}$ $ (1/\mu m^{2}) $ & $0.015$ & 0.0085 & 0.0085 \\
     $ f_{pq}$ $ (1/\mu m^{2}) $ & $0.008$ & $0.017$ & 0.0045  \\
      $ \sqrt{f_{pp}f_{qq}}/2f_{pq}$ & $1.22$ & $0.435$ & 1.25  \\
      \hline
     \end{tabular}
     \label{table_IMMI_parameter}
     \end{table}
The equation implies that, for MI process to occur, the wavenumber K of the perturbation must possesses a non-zero imaginary part and manifest itself by an exponential growth of the amplitude of the perturbation. The power gain G, which is a measure of efficiency of MI process, is defined as, $G(\Omega)=2|Im(K)|$, where K is the eigenvalue of the matrix [M] with highest imaginary part. Detailed analysis of the Eq. (\ref{eigenvector_eq}) describes the necessary condition for MI phenomena as, $\sqrt{\textit{f}_{pp}\textit{f}_{qq}}/2f_{pq} <1$ \cite{IM_MI_greadedindex_fiber}, i.e., the cross-phase modulation (XPM) term will be greater than self-phase modulation term (SPM). Table \ref{table_IMMI_parameter} shows the calculated values of IM-MI parameters for different mode combinations and indicates that among three different mode combinations, the condition to achieve IM-MI process has been satisfied for the mode group combination $LP_{01}$ and $LP_{02}$ only. For the theoretical calculations of MI gain of our experimental fiber, we have used the following parameters which have been calculated for the mode combination $LP_{01}$ and $LP_{02}$ of the CLF fiber at pump wavelength 1064 nm: $\beta_{2p} = 0.01618$ $ps^{2}$/m, $\beta_{2q} = 0.01183$ $ps^{2}$/m, $f_{pp} (1/\mu m^{2}) =  0.0257$, $f_{qq} (1/\mu m^{2}) = 0.0085$, $f_{pq} (1/\mu m^{2}) = 0.017$ and $\delta_{pq} = 0.75 ps/m $. The manifestation of IM-MI phenomena with varying peak pump power (P=Q) for the mode combination $LP_{01}$ and $LP_{02}$ is shown in Fig. \ref{gain_plot} where the gain spectra have been plotted as a function of frequency detuning $\Omega$/2$\pi$ with different peak power levels.  It is observed that IM-MI region broadens with the increase in power. The optimum modulation frequency (OMF) which is defined as the frequency at which IM-MI gain attain its maximum value, also shift towards higher value with the increase in pump power which also support our experimental findings. Finally the gain plot for varying GVM is shown in Fig. \ref{GVM_variation}, where the power is kept fixed for both the modes at 3.75 kW. It is observed that large GVM leads to higher value of the gain spectra and simultaneously the optimal modulation frequency move towards longer frequency shift. Fig. \ref{OMF_GAIN_shift} shows the variation of OMF and peak gain as a function of GVM of the participating modes. Simulated results reveal that peak gain increases gradually up to GVM 1.5 ps/m and then it tends to stabilize with further increasing GVM. 

\section{Experimental setup and results}
\begin{figure*}[hbt!]
   \centering
   \includegraphics[width=14 cm,height=5.5 cm]{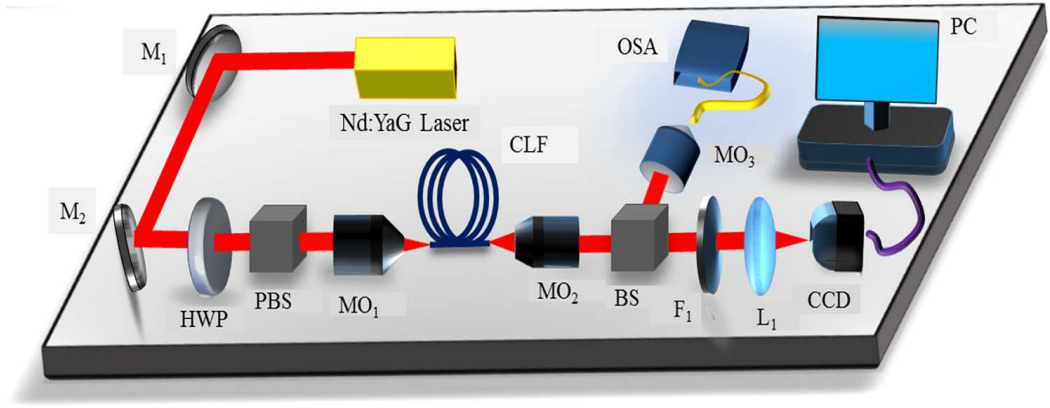}
   \caption{{\small Schematic of the experimental set-up. $M_{1},M_{2}$: silvered mirror, HWP: half wave plate, PBS: polarization beam splitter, $MO_{1},MO_{2},MO_{3}$: microscope objective, BS: plate beam splitter, $F_{1}$ : laser line filter, $L_{1}$: convex lens, CCD: charged coupled device, OSA: optical spectrum analyzer.}}
   \label{exp_setup}
   \end{figure*}

The schematic of the experimental setup is shown in Fig. \ref{exp_setup}. The pump is an output from Q-switched microchip Nd:YAG laser generating central wavelength at 1064 nm with the pulse duration of 0.77 ns. The output power of the pulses are controlled by the combination of HWP and PBS. The pump is coupled into  CL optical fiber using microscope objective (NA=0.4, 20X).
   \begin{figure}[hbt!]
      \centering
      \includegraphics[width=8 cm,height=7 cm]{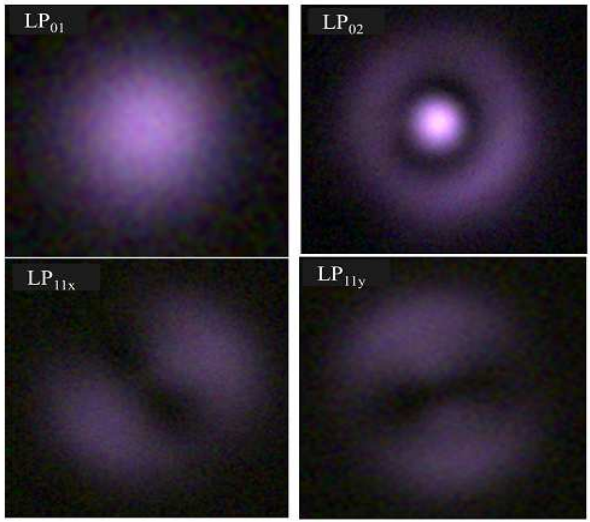}
      \caption{{\small Spatial group of mode profile experimentally identified at 1064 nm.}}
      \label{spacial_modes}
      \end{figure}
\begin{figure}[hbt!]
      \centering
      \includegraphics[width=8.5 cm,height=5 cm]{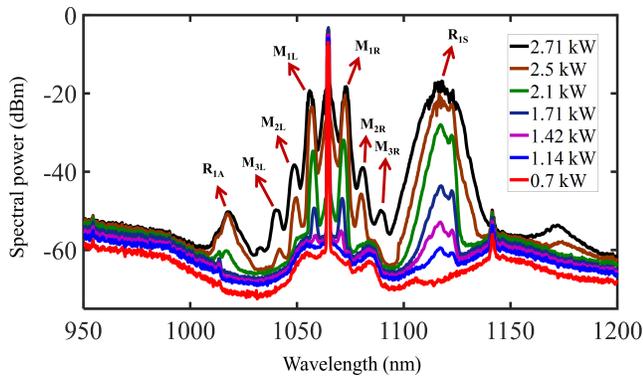}
      \caption{{\small Output spectrum for different pump powers for a 6 m long CLF.}}
      \label{power_var_6m}
      \end{figure}
\begin{figure}[hbt!]
      \centering
      \includegraphics[width=8 cm,height=5 cm]{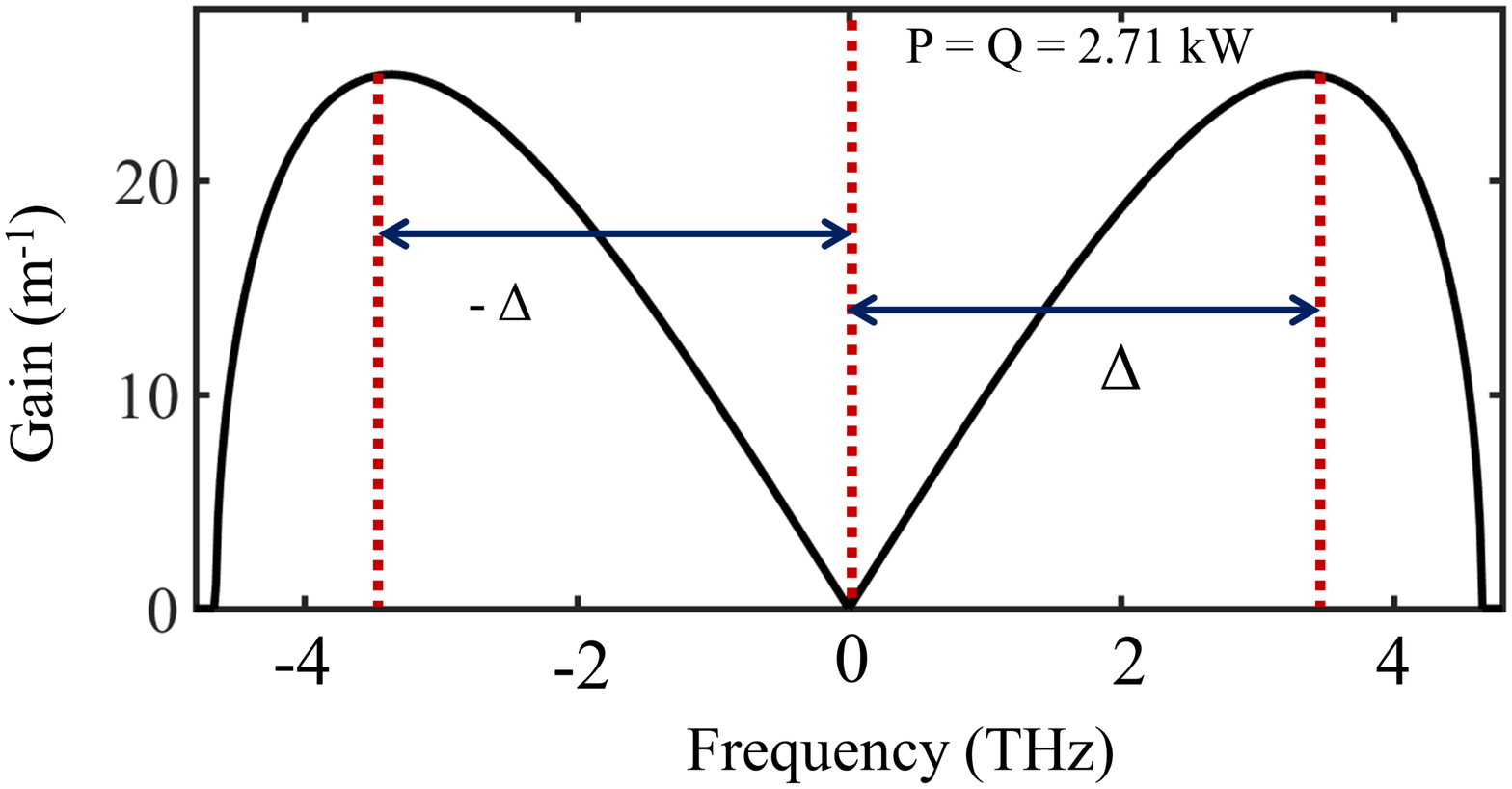}
      \caption{{\small Theoretical plot of gain spectrum as a function of frequency shift ($\Omega/2\pi$.}) considering the parameters of CLF for the peak pump power of 2.71 kW (P = Q = 2.71 kW) }
      \label{gain_2_71kw}
      \end{figure}
\begin{figure}[hbt!]
      \centering
      \includegraphics[width=8.5 cm,height=4.5 cm]{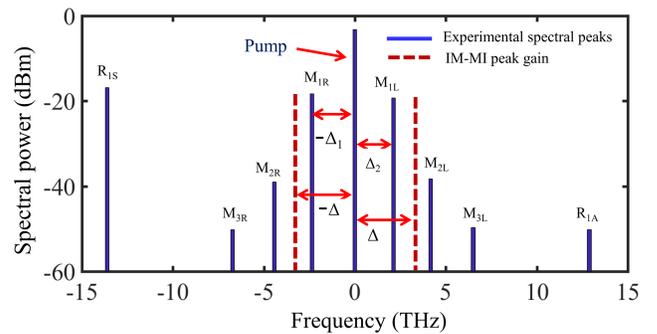}
      \caption{{\small Solid lines show the frequency shift ($\Omega/2\pi$.}) of the experimental spectral peaks at the output of 6 m long CLF for peak pump power 2.70 kW, whereas the dashed lines show the calculated theoretical positions of the peak gain of IM-MI for the same experimental conditions.}
      \label{frequency_shifting}
      \end{figure}
\begin{figure*}[t]
      \centering
      \includegraphics[width=17.5 cm,height=9.8 cm]{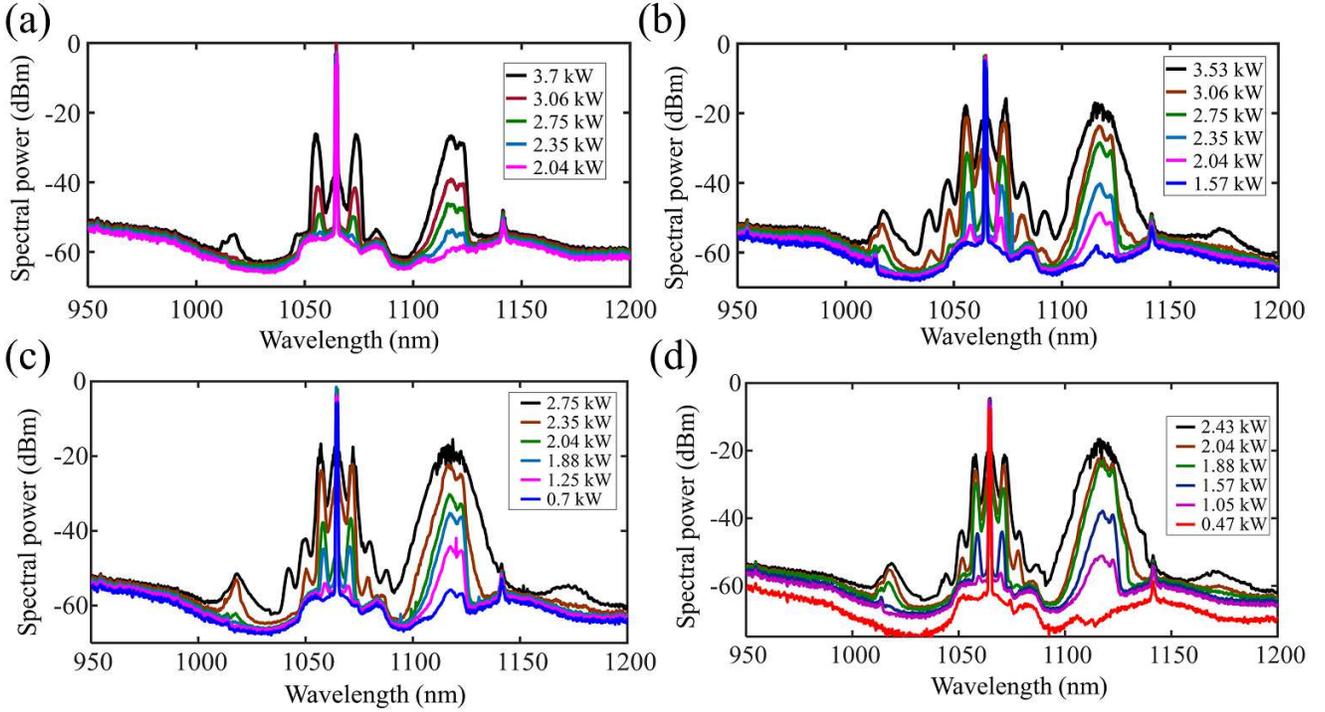}
      \caption{{\small Output spectrum for different fiber lengths (a) 4 m (b) 5m (c) 7m and (d) 8m of CLF.}}
      \label{MI_length_variation}
      \end{figure*}
 \begin{figure}[t]
      \centering
      \includegraphics[width=8 cm,height=4.5 cm]{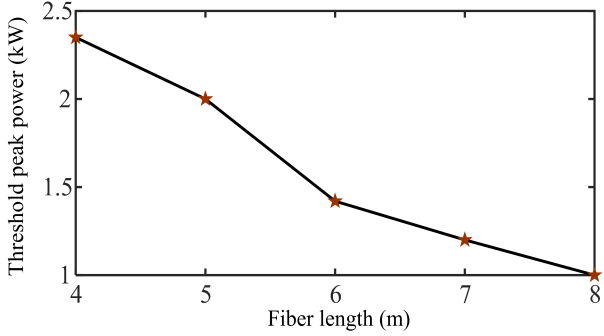}
      \caption{{\small Threshold power for IM-MI with fiber length.}}
      \label{threshold_power}
      \end{figure}
\begin{figure}[hbt!]
      \centering
      \includegraphics[width=8 cm,height=8 cm]{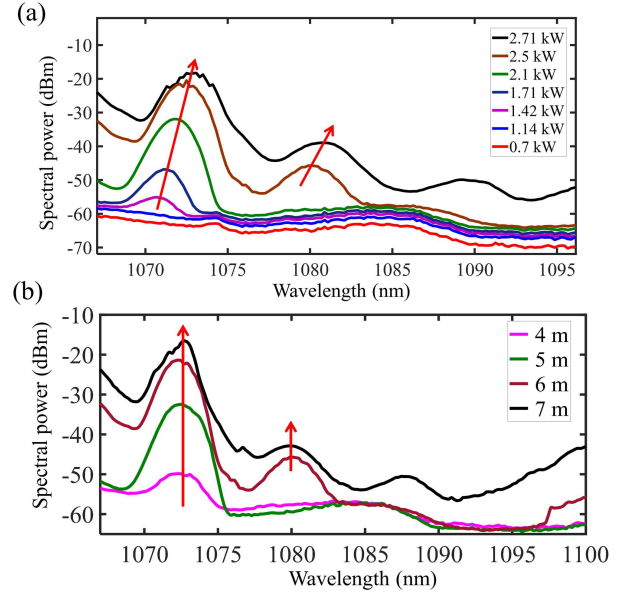}
      \caption{{\small Output spectrum for (a) various pump powers and (b) different fiber lengths. For (a) fiber length is fixed at 6 m, and for (b) peak power is fixed at 2.71 kW. MI peaks shift for the case (a) whereas no shift in wavelength is observed when the fiber length is varied.}}
      \label{wavelength_shift}
      \end{figure}
      \begin{figure}[t]
      \centering
      \includegraphics[width=8 cm,height=4.5 cm]{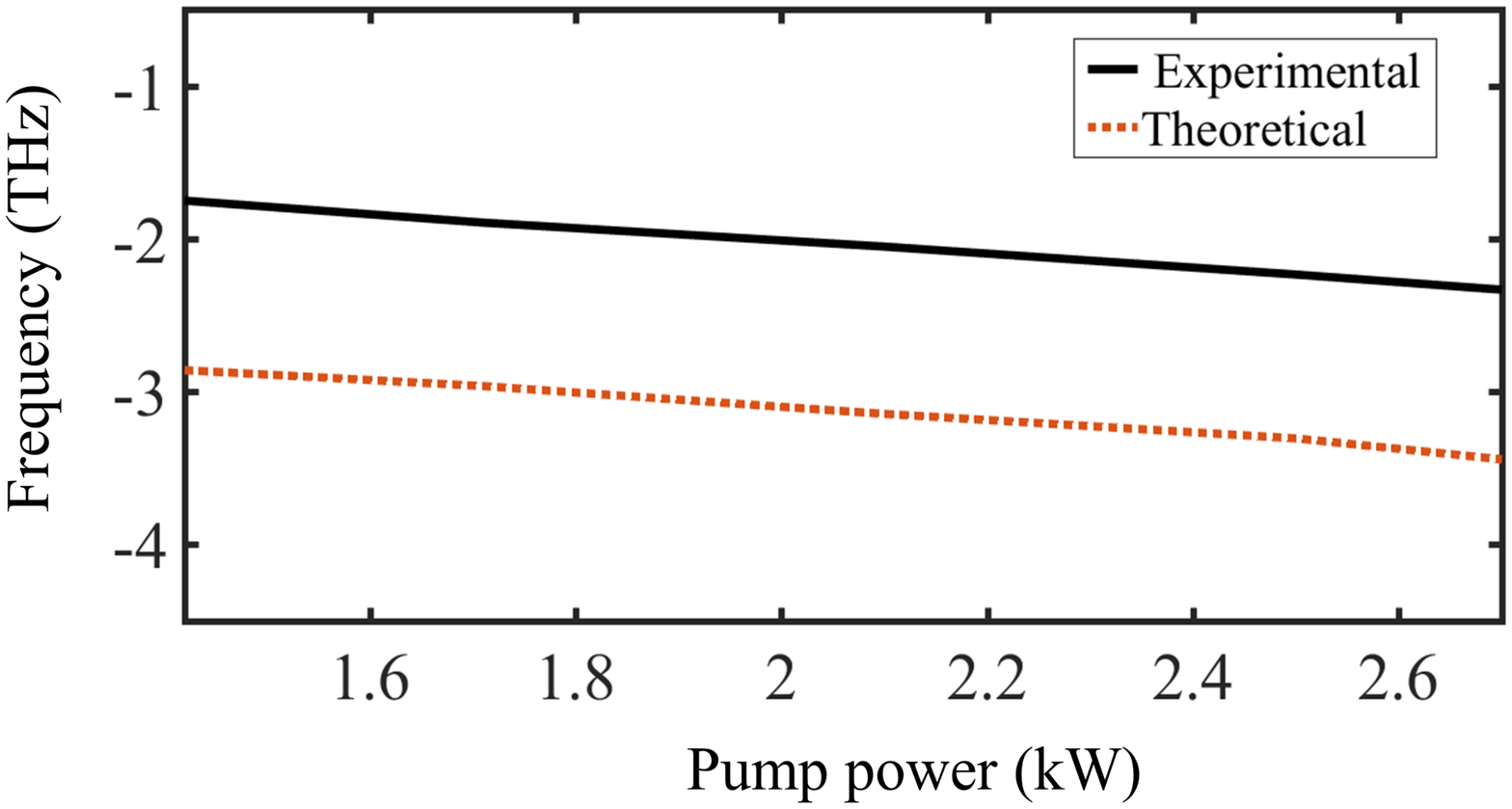}
      \caption{{\small Experimental (solid line) and theoretical (dashed line) comparison of frequency shift ($\Omega/2\pi$) of the IM-MI peaks as a function of peak pump power.}}
      \label{frequency_shift_exp_theoretical}
      \end{figure}

A three axis translational stage is used to excite selectively the desired modes and the output mode profile is detected with CCD as shown in the figure. The output spectrum is recorded by optical spectrum analyzer (OSA). All the supported spatial modes experimentally identified at 1064 nm are shown in Fig. \ref{spacial_modes} and agrees well with the simulated modal profiles as shown in Fig. \ref{modal_profile}. To study IM-MI, we have taken 6 m long CLF and pump is launched to excite the pair of modes ($LP_{01}$ and $LP_{02}$) with equal power in each modes. By finely adjusting the launching conditions of the pump pulses at the fiber input, we were able to excite equally the circularly symmetric modes. The recorded output spectrum with varying total input peak power is shown in Fig. \ref{power_var_6m}. It is observed that with the increase in power, multiple MI sidebands are generated together with the Raman Stokes and anti-Stokes peaks. For 2.71 kW of peak power, IM-MI peaks are generated at wavelengths 1073 nm ($M_{1R}$) and 1056 nm ($M_{1L}$). Harmonics are generated in the red-side of the pump nearly at  1081 nm ($M_{2R}$) and 1090 nm ($M_{3R}$) whereas, on the blue-side of the pump at 1048.5 nm ($M_{2L}$) and 1040 nm ($M_{3L}$) wavelengths. Raman Stokes and anti-Stokes are generated at 1118 nm ($R_{1}$) and 1017.5 nm ($R_{2}$), respectively. The IM-MI gain for the pump power of 2.71 kW in each modes are shown in Fig. \ref{gain_2_71kw} which yields that maximum gain occurs at the frequency shift of $\pm \Delta$, where $\Delta = 3.45$ THz. The position of the spectral peaks generated through 6 m fiber length with 2.71 kW pump power, are shown in Fig. \ref{frequency_shifting}  as a function of frequency shift from the pump wavelength. The IM-MI peaks are generated at $\Delta_{1} = - 2.36$ THz and $\Delta_{2} = 2.14$ THz apart. The asymmetry in frequency shift occurs due to the effect of higher order dispersion coefficient. Considering up to second-order dispersion coefficient, the position of the peaks of theoretical IM-MI gain are indicated by the dashed lines. The slight difference of the experimental and the theoretical observations are due to the refractive index profile that used in COMSOL for calculating dispersion parameters has been fitted and extrapolated at the pump wavelength (1064nm) from the original profile. To investigate the influence of fiber length, the evolution of IM-MI spectra for different fiber lengths with varying peak pump powers are measured and shown in Fig. \ref{MI_length_variation}. It is observed that, the threshold power require to build up spectral IM-MI peaks from noise gradually decreases with the increase in fiber length and shown in Fig. \ref{threshold_power}.  Also, Raman threshold power reduces with the increase in fiber length. Longer fiber length provides effective platform to break-up the input pulses into multiple peaks through IM-MI even with sufficient low input pump power. Efficient IM-MI peaks with strong Stokes and anti-Stokes wave are generated with very low input pump power (2.43 kW in each mode) using 8m long CLF which is shown in Fig. \ref{MI_length_variation} (d). The shift of spectral peaks with varying pump powers and fiber lengths is shown in Fig. \ref{wavelength_shift}(a) and \ref{wavelength_shift}(b), respectively. The fiber length has been fixed to 6 m. It is clearly evident that the IM-MI peaks and the cascaded harmonics exhibit shift in wavelength, as depicted in Fig. \ref{wavelength_shift}(a). The experimental observation of OMF shift of the IM-MI peaks as a function of peak pump power is shown by the solid line in Fig. \ref{frequency_shift_exp_theoretical}, whereas the dashed line shows the theoretical prediction. The theoretical and the experimental lines shows almost same slope. It is important to note that MI peaks do not shift with the variation in fiber length, as shown in Fig. \ref{wavelength_shift}(b) where input peak pump power is fixed at 2.71 kW.

\section{Conclusion}
To conclude our work, we have reported detail theoretical and experimental observation of IM-MI in step-index scenario which yields that step-index multimode fiber can provide excellent platform for the realization of IM-MI. The modal GVM plays an important role in the gain spectrum and large GVM provides strong IM-MI peaks. Furthermore, details investigation with varying pump power and fiber length is demonstrated with unified manner. We also demonstrate the shifting of optimum frequency with pump power. The study can be effectively extended by employing large core step-index MMF which can support many higher order modes and launching the pump in different wavelength with non-identical spatial modes providing large GVM. The GVM can also be tailored by choosing the participating modes. Our observation will pave the way in multitude of applications such as the realization of coherent wideband frequency generation, formation of highly repetition rate vector soliton trains and so on.

\end{document}